\documentclass[]{spie}  

\usepackage{amsmath,amsfonts,amssymb}
\usepackage{graphicx}
\usepackage[colorlinks=true, allcolors=blue]{hyperref}
\usepackage{gensymb}
\usepackage [english]{babel}
\usepackage [autostyle, english = american]{csquotes}
\MakeOuterQuote{"}

\title{Improve Global Glomerulosclerosis Classification with Imbalanced Data using CircleMix Augmentation}

\author[a]{Yuzhe Lu}
\author[b]{Haichun Yang}
\author[a]{Zheyu Zhu}
\author[a]{Ruining Deng}
\author[b]{Agnes B. Fogo}
\author[a]{Yuankai Huo}

\affil[a]{Department of Electrical Engineering \& Computer Science, Vanderbilt University, Nashville, TN, USA 37235}
\affil[b]{Department of Pathology, Microbiology \& Immunology, Vanderbilt University Medical Center, Nashville, TN, USA 37232}

\authorinfo{Further author information: (Send correspondence to Yuankai Huo)\\Yuankai Huo: E-mail: yuankai.huo@vanderbilt.edu}

\begin{document} 
\maketitle

\begin{abstract}
The classification of glomerular lesions is a routine and essential task in renal pathology. Recently, machine learning approaches, especially deep learning algorithms, have been used to perform computer-aided lesion characterization of glomeruli. However, one major challenge of developing such methods is the naturally imbalanced distribution of different lesions. In this paper, we propose CircleMix, a novel data augmentation technique, to improve the accuracy of classifying globally sclerotic glomeruli with a hierarchical learning strategy. Different from the recently proposed CutMix method, the CircleMix augmentation is optimized for the ball-shaped biomedical objects, such as glomeruli. 6,861 glomeruli with five classes (normal, periglomerular fibrosis, obsolescent glomerulosclerosis, solidified glomerulosclerosis, and disappearing glomerulosclerosis) were employed to develop and evaluate the proposed methods. From five-fold cross-validation, the proposed CircleMix augmentation achieved superior performance (Balanced Accuracy$=73.0\%$) compared with the EfficientNet-B0 baseline (Balanced Accuracy$=69.4\%$).
\end{abstract}


\section{INTRODUCTION}
\label{sec:intro}  

The identification of non-sclerotic and sclerotic glomeruli is essential to compute percentage of global glomerulosclerosis, a quantitative measurement corresponding to several critical clinical outcomes. With fine-grained definition, globally sclerotic glomeruli, also called global glomerulosclerosis, can be characterized into three categories: obsolescent glomerulosclerosis, solidified glomerulosclerosis, or disappearing glomerulosclerosis\cite{marcantoni2002hypertensive}. As globally sclerotic glomeruli occur with both aging and kidney diseases, the fine-grained phenotype would provide more precise evidence to support both scientific research and clinical decision making. However, differentiating these patterns typically requires heavy manual efforts by trained clinical experts, which is not only tedious, but also labor-intensive.
Therefore, there is a strong need to develop automatic classification algorithms to perform fine-grained glomerulosclerosis classification, especially with an increasingly large amount of digitized data from whole slide imaging (WSI). 

In the past few years, many studies have been conducted to classify different glomerular lesions using computer-aided approaches~\cite{marsh2018deep,zeng2020identification,uchino2020classification,ginley2019computational,ginley2020fully}. However, there are few, if any, studies that have developed deep learning approaches for fine-grained classification of glomerular lesions to characterize the globally sclerotic glomeruli into three categories: obsolescent glomerulosclerosis, solidified glomerulosclerosis, and disappearing glomerulosclerosis. Such fine-grained characterization is challenging, as the available data are typically highly imbalanced. For instance, the prevalence of obsolescent glomerulosclerosis is naturally much higher than solidified or disappearing glomerulosclerosis, leading to the technical difficulty which is well known as the “imbalanced classes problem”.  

In this paper, we propose CircleMix, a novel data augmentation technique, to improve the accuracy for classifying non-sclerotic and sclerotic glomeruli, as well as fine-grained sub-types of globally sclerotic glomeruli. Our CircleMix algorithm is inspired by the prevalent CutMix augmentation~\cite{yun2019cutmix}, yet is optimized for ball-shaped biomedical objects, such as glomeruli in this study (Figure 1). To further enhance the performance of imbalanced classes, the training is modeled as a hierarchical training procedure. To train and evaluate the deep learning algorithms, we collected images from 6,861 glomeruli with five classes (normal, periglomerular fibrosis, obsolescent glomerulosclerosis, solidified glomerulosclerosis, and disappearing glomerulosclerosis) 

\begin{figure}[t]
\begin{center}
\includegraphics[width=0.9\textwidth]{{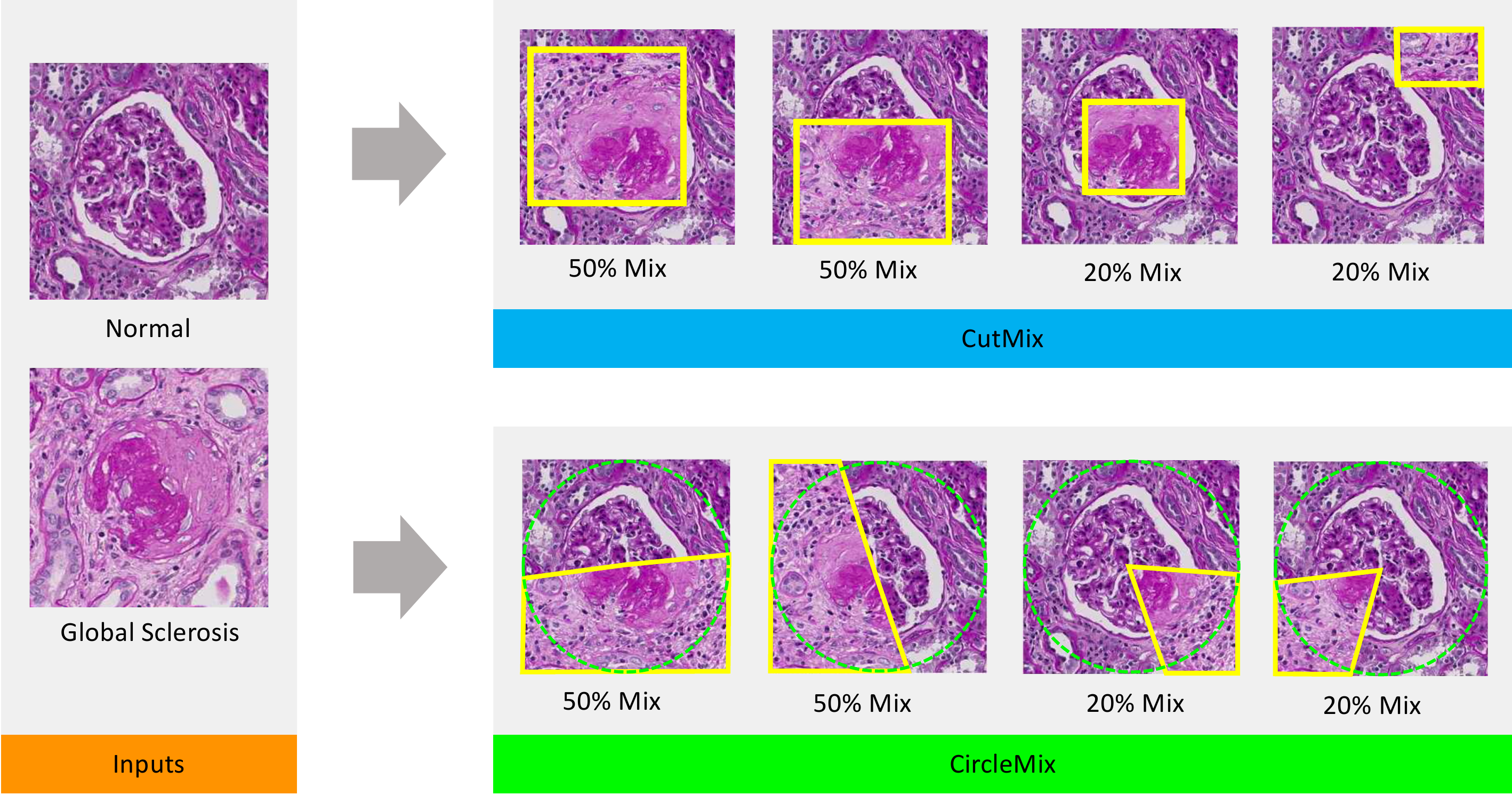}}
\end{center}
\caption{This figure shows the examples of performing different data augmentation strategies. The left panel shows the examples of glomerular image patches, which can be achieved from either object detection~\cite{yang2020circlenet} or manual annotation~\cite{2020arXiv200713952Z}. The glomeruli are typically located in the central location within the image patches. In the upper right panel, the morphological features from one glomerulus can be largely lost when performing CutMix. By contrast, the CircleMix maintains the morphological features from both glomeruli with different percentages of mixture.} 
\label{Fig.1} 
\end{figure}

To summarize, the contribution of this work is three-fold: 
\begin{itemize}
	\item We proposed the CircleMix, a novel data augmentation algorithm that is optimized for ball-shaped biomedical objects.
	\item We evaluated the performance of hierarchical learning strategy on the fine-grained classification of glomeruli with imbalanced data distribution.
	\item To the best of our knowledge, this is the biggest study so far (6,861 glomeruli) to investigate deep learning based image classification on both (1) non-sclerotic vs. sclerotic glomeruli, and (2) fine-grained classification of obsolescent, solidified, and disappearing glomerulosclerosis.
	
\end{itemize}

\section{Related Work}

The groundbreaking learning capability provided by deep learning algorithms comes largely from the unprecedented large number of parameters in neural networks. To improve the generalizability of the deep neural networks, applying data augmentations is typically an inevitable step to introduce extra randomness to the training data~\cite{shorten2019survey}. Beyond the standard single image based augmentation strategies, Yun et al.~\cite{yun2019cutmix} proposed a novel data augmentation strategy, which is called CutMix, by mixing different images as new training data. By cutting and pasting patches among training images, CutMix forced the deep networks to provide partial decisions on a mixed image, which achieved the superior performance compared with benchmarks (e.g., Mixup ~\cite{zhang2017mixup}). However, one major problem of CutMix is that the random patch-based image fusion might lose discriminative features from the source images. To optimize the mixing procedure for ball-shaped biomedical objects, the novel CircleMix algorithm is proposed in this study.

In image classification, there is a long-lasting issue called imbalanced classes problem. The problem occurs when the numbers of samples are considerably imbalanced (e.g., one class can have ten times or more samples than another), where the predictions from the trained neural networks are typically biased to the majority class. Many previous efforts have been made to improve the performance on imbalanced data, such as data sampling\cite{chawla2002smote}, cost-sensitive learning\cite{ling2008cost}, and their combination\cite{tang2008svms, huang2016learning}. In this study, we explored the effect of hierarchical learning strategy to perform fine-grained classification on an imbalanced glomerular cohort.

\section{Methods}
\label{sec:sections}

\subsection{CircleMix}
In this paper, we propose CircleMix, a novel data augmentation technique optimized for ball-shaped glomeruli (Figure 1). Firstly, the start and end points on the sides of the image are randomly generated. Then, together with the image center and corners between the start and end points, a polygon mask is produced, which is then filled with the corresponding pixels from the other training image. 
We define $I \in \mathbb{R}^{H \times W \times C}$ as an input image with $H \times W$ resolution and $C$ channels (e.g., three channels for RGB). $Y$ is the one-hot-vector label of class for image $I$. By performing CircleMix, a new training sample $(\Tilde{I},\Tilde{Y})$ is formed by combining two training images $(I_{A}, Y_{A})$ and $(I_{B}, Y_{B})$. 
The procedure is presented as the following equations
\begin{equation}
\begin{split}
    \Tilde{I} & =  \mathbf{M} \odot I_{A} + (\mathbf{1}- \mathbf{M}) \odot I_{B} \\
    \Tilde{Y} & =  \lambda Y_A + (1-\lambda) Y_B, \\
\end{split}
\label{eq:circlemix}
\end{equation}
where $\mathbf{M} \in \{0,1\}^{H \times W}$ is a polygon mask for filling image $A$, while $(\mathbf{1}- \mathbf{M})$ is the remaining polygon region for filling image $B$. ``$\odot$'' is element-wise multiplication. $\lambda$ is calculated by $(r_{end} - r_{start})/ 360$, where we sample the start and the end points of the mask, which correspond to unique points on the sides of the training image, as the following:

\begin{equation}
    r_{start}, r_{end} \sim \text{Uniform}~ \left(0, 360\right), ~~~r_{start}, r_{end}  = min(r_{start}, r_{end}), max(r_{start}, r_{end})
\label{eq:mix}
\end{equation}

In implementation, the CircleMix is performed by randomly combining two training samples within the same mini batch, according to Eq.\ref{eq:circlemix}.

\subsection{Hierarchical Learning}
Our proposed training framework is defined as a hierarchical architecture, as shown in Figure 2. 

Concretely, we trained a five-class classifier first. Then, we used the best five-class model from validation to fine-tune three children classifiers, with one to re-verify the classification of normal and periglomercular fibrosis, one to re-verify the classification of three global glomerulosclerosis types, and one to re-verify global solidified and global disappearing types. With each of these children classifier, we combined its predictions with that of the five-class classifier to produce the final results. Specifically, if prediction from the parent classifier falls into the set of classes the child classifier is re-verifying, the final prediction will be decided by the child classifier.

\begin{figure}[t]
\begin{center}
\includegraphics[width=0.9\textwidth]{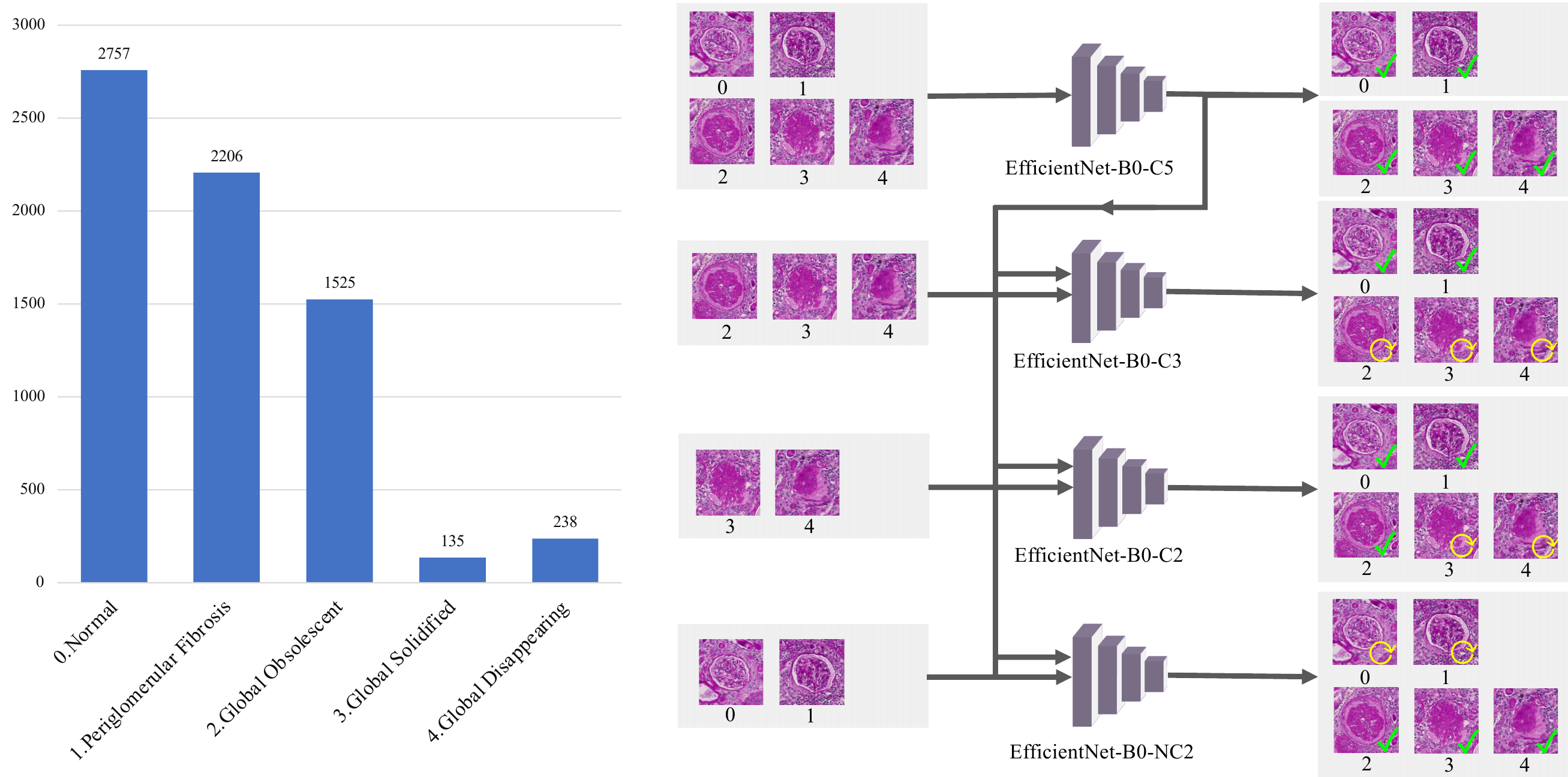}
\end{center}
\caption{This figure shows the hierarchical learning framework. The left panel shows the imbalanced data distribution of our data. The right panel shows the hierarchical design. EfficientNet-B0-C5 is used to classify all five classes, and then used to fine-tune children classifiers. Specifically, EfficientNet-B0-C3 is fine-tuned to perform classification on classes ``2'', ``3'', and ``4'', EfficientNet-B0-C2 is fine-tuned to perform classification on classes ``3'' and``4'', and EfficientNet-B0-NC2 is fine-tuned to perform classification on classes ``0'' and ``1''. } 
\label{Fig.2} 
\end{figure}

\section{Experiments and Results}

\subsection{Data}
The human nephrectomy tissues were acquired from 23 patients, whose tissues were routinely processed and paraffin-embedded, with 3 $\mu$m thickness sections cut and stained with PAS. 6,861 glomeruli were extracted from WSI using the EasierPath semi-manual annotation software~\cite{2020arXiv200713952Z}. Then, all glomeruli were manually labeled, including 2,757 normal glomeruli, 2,206 periglomerular fibrosis glomeruli, 1,525 global obsolescent glomeruli, 135 global solidified glomeruli, and 238 global disappearing glomeruli. The images were resized to 256$\times$256-pixel for training, validation, and testing. The data were de-identified, and studies were approved by the Institutional Review Board (IRB).

\subsection{Experiment Design}
In the experiments, EfficientNet-B0~\cite{tan2019efficientnet} is employed as the backbone model of classification due to its high efficiency of learning large-scale images. We adapted the EfficientNet-B0 model pretrained on ImageNet\cite{deng2009imagenet} by customizing the fully-connected layers based on our tasks. The model was trained and tested with standard five-fold cross-validation. Briefly, the data was split into five folds at subject level, where each fold was withheld as testing data once. The remaining data for each fold was split as 75\% training data and 25\% validation data. Therefore, for each fold, the final split was 60\% training, 20\% validation, and 20\% test. To avoid data contamination, all glomeruli from a patient were used either for training, validation or testing. 

The model was trained using cross entropy loss with stochastic gradient descent optimizer and a batch size of 16. We started with a learning rate of 0.01 and decayed it by 10 half way through the total number of training epochs. We used both balanced accuracy and balanced $F1$ score to evaluate the performance. The balanced accuracy was calculated as the arithmetic mean of each class's recall, and balanced $F1$ was calculated as the arithmetic mean of each class's $F1$ score (one vs. remaining). In each fold, the model with the best balanced accuracy in the validation set was selected for testing. 

\subsection{Basic Augmentation}
The basic data augmentations we used include horizontally and vertically flipping $50\%$ of all training images and randomly cropping $0-16$ pixels. They are applied to all the experiments in this study.

\subsection{Results}
\subsubsection{Non-hierarchical Training}
We evaluated the performance of CircleMix by training EfficientNet-B0~\cite{tan2019efficientnet} as (1) a standard binary classifier, and (2) a five-class classifier, without performing the hierarchical training. The binary classifier ("Binary" in Table 1) classified all images as two classes: global glomerulosclerosis or others. The five-class classifier ("Five-class" in Table 1) performed the fine-grained five class classification. 

\begin{figure}
\begin{center}
\includegraphics[width=0.9\textwidth]{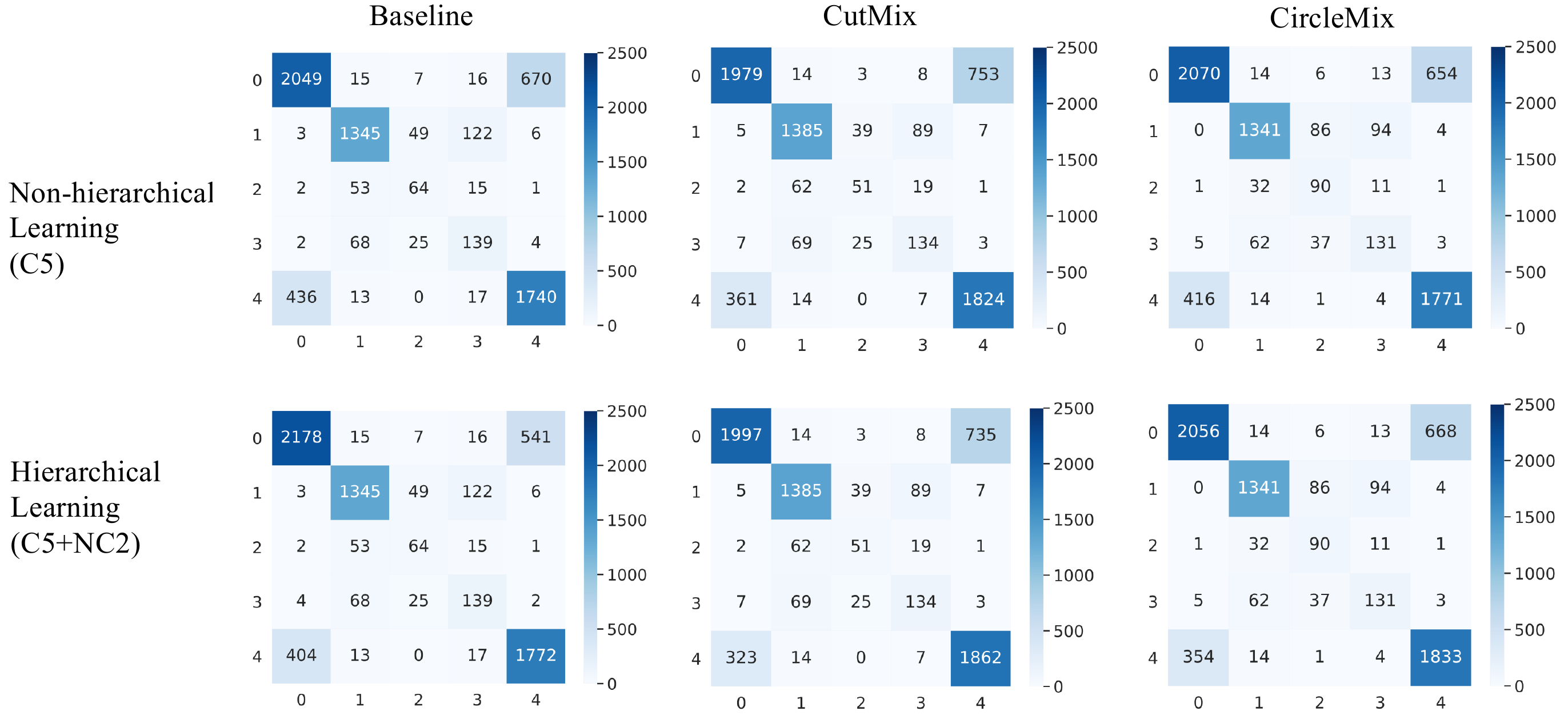}
\end{center}
\caption{This figure shows the detailed confusion matrix of different data augmentation and learning strategies.} 
\label{Fig.3} 
\end{figure}

From the results, when applied the proposed CircleMix augmentation, the model achieved superior performance on both binary classification and five-class classification tasks in terms of balanced accuracy (ACC) and balanced $F1$ score compared to the baseline model and the model using CutMix augmentation. The binary classifiers to differentiate non-sclerotic and sclerotic glomeruli perform extremely well across different settings, improving more than $7\%$ in $F1$ score compared to previous studies\cite{marsh2018deep} thanks to a much larger dataset. For the five-class classification task, CircleMix helps to improve the balanced accuracy by over $3\%$ and balanced $F1$ by $2\%$ against the baseline. From the confusion matrix in Figure \ref{Fig.3}, this improvement mainly comes from the the better classification of the least populated class. By contrast, CutMix predicts the least populated class worse than the baseline, which results in a significant degredation of its overall performance.

\begin{table}
\caption{Non-hierarchical Training.}\label{tab1}
\centering
{%
\begin{tabular}{l  c c c c} 
\hline
 &\multicolumn{2}{c}{Binary} & \multicolumn{2}{c}{Five-class}\\
\hline
 & ACC & F1 & ACC & F1\\
\hline
EfficientNet-B0\cite{tan2019efficientnet} & 98.9\% & 98.0\% & 69.4\% & 67.8\%\\ 
EfficientNet-B0 + CutMix\cite{yun2019cutmix} & 99.1\% & 98.4\% & 67.9\% & 67.7\% \\
EfficientNet-B0 + CircleMix (ours)& \textbf{99.2}\% & \textbf{98.6}\% & \textbf{73.0}\% & \textbf{69.8}\%\\  
\hline
\end{tabular}
}
\\
* ``Binary'' is the binary classification results of global glomerulosclerosis vs. others, while ``Five-class'' is the fine-grained five class classification. ``ACC'' is the balanced accuracy score. ``F1'' is the balanced $F1$ score.
\break
\\
\end{table}

\begin{table}
\caption{Hierarchical Training.}\label{tab2}
{%
\begin{tabular}{l c c c c c c c c } 
\hline
&\multicolumn{2}{c}{C5} & 
\multicolumn{2}{c}{C5+NC2} & 
\multicolumn{2}{c}{C5+C3} & \multicolumn{2}{c}{C5+C2}\\
\hline
& ACC & F1 & ACC & F1 & ACC & F1 & ACC & F1\\ 
\hline
EfficientNet-B0\cite{tan2019efficientnet}  &69.4\% & 67.8\% & 70.7\% & 69.1 \% & \textbf{67.9}\% & 67.8\%
& 68.6\% & 67.0\%\\ 
EfficientNet-B0+CutMix\cite{yun2019cutmix} 
& 67.9\% & 67.7\% & 68.3\% & 68.1\%  & 62.6\% & 64.4\% 
& 67.9\% & 67.6\%\\ 
EfficientNet-B0+CircleMix (ours) 
& \textbf{73.0}\% & \textbf{69.8\%} & \textbf{73.5}\% & \textbf{70.2}\%  & 66.7\% & \textbf{68.4}\% & \textbf{71.8\%} & \textbf{69.0\%}\\ 
\hline
\end{tabular}
}
\\
* ``NC2'', ``C2'', ``C3'', and ``C5'' represent EfficientNet-B0-NC2, EfficientNet-B0-C2, EfficientNet-B0-C3, and EfficientNet-B0-C5, respectively. ``C5+X'' indicates the merged results using EfficientNet-B0-C5 and EfficientNet-B0-X.\\
\end{table}

\subsubsection{Hierarchical Training}
Next, we evaluated the performance of hierarchical training with different hierarchical combinations (Table 2). The ``C5'', ``C3'', ``C2'', and ``NC2'' represented the four deep networks in Figure 2. 

Based on the experimental results, while EfficientNet-B0-C5 and EfficientNet-B0-NC2 together produces slightly better results, other combinations generally give inferior performance. We observed a performance degredation of the ``C3''and ``C2'' classifiers compared to the ``C5'' classifier. This might be because EfficientNet-B0-C3 and EfficientNet-B0-C2 are trained with too few data points due to the imbalanced nature of the dataset. The confusion matrices from the combination of EfficientNet-B0-C5 and EfficientNet-B0-NC2 are presented together with those from the non-hierarchical experiments in Figure 3.

\section{Conclusions}  
In this paper, we proposed CircleMix, a novel data augmentation algorithm optimized for ball-shaped biomedical image classification, which is able to outperform the baseline and the state-of-the-art CutMix augmentation in glomerular classification task. To address the imbalanced classes problem, we evaluated the performance of the hierarchical training strategy on the fine-grained glomerular classification task. Though this strategy shows mixed results, the best overall performance was nonetheless achieved by combining the CircleMix augmentation with hierarchical training, compared with other experiments.


\section{ACKNOWLEDGMENTS}       
 
This work was supported by NIH NIDDK DK56942(ABF).  This work has not been submitted for publication or
presentation elsewhere. 


\bibliographystyle{spiebib} 
\bibliography{paper}

\end{document}